# HIGH AND LOW-TEMPERATURE CRYSTAL AND MAGNETIC STRUCTURES OF ε-Fe$_2$O$_3$ AND THEIR CORRELATION TO ITS MAGNETIC PROPERTIES

M. Gich[a], C. Frontera, A. Roig, E. Taboada, E. Molins

*Institut de Ciència de Materials de Barcelona, Consejo Superior de Investigaciones Científicas, Campus de la UAB, 08193 Bellaterra, Catalunya, Spain*

H. R. Rechenberg

*Instituto de Física, Universidade de São Paulo, Caixa Postal 66318, 05315-970 São Paulo, Brazil*

J. D. Ardisson, W. A. A. Macedo

*Laboratório de Física Aplicada, Centro de Desenvolvimento da Tecnologia Nuclear, 31270-901 Belo Horizonte, Brazil*

C. Ritter

*Institut Laue-Langevin, 6 rue Jules Horowitz , BP 156, 38042 Grenoble Cedex, France*

V. Hardy

*Laboratoire CRISMAT/ENSICAEN, UMR 6508 du CNRS,6 Bd Marechal Juin, 14050 Caen, France.*

J. Sort, V. Skumryev, J. Nogués.

*Institució Catalana de Recerca i Estudis Avançats (ICREA) and Departament de Física, Universitat Autònoma de Barcelona, 08193 Bellaterra, Catalunya, Spain*

[a]Email: mgich@icmab.es

**Abstract**

The crystal and magnetic structures of the orthorhombic ε-Fe$_2$O$_3$ have been studied by simultaneous Rietveld refinement of X-ray and neutron powder diffraction data in combination with Mössbauer spectroscopy, as well as magnetisation and heat capacity measurements. It has been found that above 150 K the ε-Fe$_2$O$_3$ polymorph is a collinear ferrimagnet with the magnetic moments directed along the *a* axis, while the magnetic ordering below 80 K is characterised by a square-wave incommensurate structure. The transformation between these two states is a second order phase transition and involves subtle structural changes mostly affecting the coordination of the tetrahedral and one of the octahedral Fe sites. The temperature dependence of the ε-Fe$_2$O$_3$ magnetic properties is discussed in the light of these results.

## I. INTRODUCTION

Despite being known since 1934,[1] $\varepsilon$-$Fe_2O_3$ has been much less studied than other iron (III) oxides such as the α and γ polymorphs. For instance, it is not since quite recently that an agreement has been reached on describing its crystalline structure with an orthorhombic cell in the *Pna2$_1$* space group.[2] The structure is isomorphous to $GaFeO_3$ and $AlFeO_3$ and presents four different Fe sites, three in octahedral and one in tetrahedral oxygen coordination. From the magnetic point of view, $\varepsilon$-$Fe_2O_3$ has been long known to be paramagnetic above 510 K and to present, below this temperature, an antiparallel arrangement of the $Fe^{3+}$ cations,[2-5] characterized by a rather large anisotropy.[5] However, no in-depth studies have been carried out up to now to determine its magnetic structure. The interest in the magnetic properties of this material has revived since the scenario disclosed by recent studies is unexpectedly rich and complex. Indeed, this oxide exhibits a huge room-temperature coercivity of about 20 kOe[6,7] and a magnetic transition at low temperature.[8,9] This low-temperature transition leads to about 5 % diminution in the saturation magnetization, $M_s$, and a reduction of the squareness ratio, $M_r/M_s$, where $M_r$ is the remanent magnetisation, from 0.5 to 0.2 on cooling from 150 to 85 K. Moreover, a large decrease in the coercivity, $H_c$, from 22.5 to 0.8 kOe is observed between 200 and 100 K[9] (see Fig. 1). In addition, the material presents a magnetoelectric coupling at about 100 K.[10]

Tronc *et al.*, studied the $\varepsilon$-$Fe_2O_3$ magnetic structure at 9 K by means of in-field Mössbauer spectroscopy and proposed a nearly collinear ferrimagnetic order at the octahedral sites and a misalignment with possible disorder at the tetrahedral sites.[2] However, the presence of a magnetic transition, which is not observed in the $GaFeO_3$ and $AlFeO_3$ ferrimagnetic isomorphous systems,[11,12] was somehow overlooked in their work and the large thermal dependence presented by the hyperfine field of the Fe tetrahedral site was ascribed to dynamical phenomena. The two groups which first reported on this low-temperature transition have interpreted it quite differently.[8,9] On the one hand, according to Kurmoo *et al.*, all $Fe^{3+}$ cations in $\varepsilon$-$Fe_2O_3$ carry the same magnetic moment and its room-temperature magnetic structure is that of a canted antiferromagnet which, on cooling to below 150 K, would undergo a Morin-like transition resulting in a second canted antiferromagnetic phase with smaller canting angles.[8] A similar interpretation to the transition has also been given by Sakurai *et. al* who also studied the $\varepsilon$-$Fe_2O_3$ structure by X-ray diffraction between 60-293 K temperature range and although they were not able to reveal any structural change along the transition.[13] In contrast, we have proposed that at room-temperature, $\varepsilon$-$Fe_2O_3$ is a collinear ferrimagnet with the net magnetization arising from the lower magnetic moment of the $Fe^{3+}$ in tetrahedral coordination and we have related the anomalies in the magnetic properties at low temperatures to the appearance of an incommensurate magnetic order, already indicated by preliminary neutron diffraction experiments.[9] In the context above described, besides clarifying the nature of the $\varepsilon$-$Fe_2O_3$ magnetic structure below 100 K, a second question that needs to be investigated is the eventual occurrence of structural changes associated with the transition. Indeed, understanding the relationship between the structural and magnetic properties of $\varepsilon$-$Fe_2O_3$ can be crucial to comprehend both the magnetic softening[9] and the coupling between magnetic and dielectric properties[10] that are observed in this oxide.

Thus, the aim of this work is to elucidate the high temperature (HT) and low temperature (LT) structures of $\varepsilon$-$Fe_2O_3$ and to shed some light into the nature of this transition. We present here a refinement of the crystal and magnetic structures of $\varepsilon$-

Fe$_2$O$_3$ from X-ray and neutron powder diffraction data which is complemented by Mössbauer spectroscopy, magnetization and heat capacity measurements.

## II. EXPERIMENTAL DETAILS

For the present study, we have used the ε-Fe$_2$O$_3$/amorphous-SiO$_2$ nanocomposite sample that was already investigated in our previous work.[9] For the powder diffraction measurements, new batches were prepared following the same procedure that has been described in detail elsewhere.[7] In some of these samples small amounts (<10 wt. %) of α-Fe$_2$O$_3$ were detected. To optimize the material structural characterization using diffraction, the SiO$_2$ amorphous matrix was removed by continuously stirring the sample in a concentrated NaOH solution (~12 M) at 80 ºC. After two days, the solution was centrifuged and the sample was washed several times with distilled water until achieving a neutral pH and finally dried at 60 ºC. X-ray diffraction analysis performed after this treatment revealed the complete removal of the SiO$_2$ amorphous matrix while the ε-Fe$_2$O$_3$ phase remained unaltered. Neutron powder diffraction (NPD) patterns were collected in the D20 instrument at the Institut Laue Langevin (Grenoble, France) using a 2.42 Å wavelength on heating at 1.7 K/min between 10 and 300 K. Synchrotron x-ray powder diffraction (SXRPD) measurements were performed at fixed temperatures on heating in the same temperature range using the ID31 (λ=0.500 Å) diffractometer of ESRF (Grenoble, France) with a 0.500 Å wavelength. The patterns were refined by the Rietveld method using the *FullProf*[14] program suite. Mössbauer spectra were acquired at different temperatures between 10 and 300 K using a conventional transmission Mössbauer spectrometer with a $^{57}$Co/Rh source. Velocity calibration was done using a 6 μm foil of metallic iron, and the Mössbauer parameters are given relative to this standard at room temperature. The program package NORMOS [15] was used to fit the spectra. Magnetization measurements were performed using a Quantum Design MPMS SQUID magnetometer. Measurements of the ac-susceptibility (1 Oe, 33-1000 Hz) and heat capacity $C_p$, using a $2\tau$ relaxation method, were carried out in a Quantum Design PPMS.

### III. RESULTS AND DISCUSSION

#### A. Neutron and X-ray diffraction

In Fig. 2, the NPD patterns of $\varepsilon$-$Fe_2O_3$ recorded at 200, 160, 80 and 10 K evidence a change of magnetic structure taking place between 160 and 80 K which is characterized by the disappearance of the (011), (110), (111) and (120) magnetic reflections and the development of satellites at either sides of these reflections. The changes observed in the NPD patterns of Fig. 2 only affect some of the magnetic reflections, suggesting that the transformation does not involve major structural changes. Whereas the HT phase of $\varepsilon$-$Fe_2O_3$ (200 and 160 K patterns) can be refined using the magnetic and nuclear structures of $AlFeO_3$ as a starting model,[12] the new peaks that characterize the magnetic structure of the LT phase can be indexed with an incommensurate propagation vector ***k***=(0, 0.1047, 0).

The small crystallite size of the $\varepsilon$-$Fe_2O_3$ nanoparticles (below 20 nm) causes a peak broadening and therefore a strong peak overlap which reduces the number of effective reflections. A multipattern approach using both NPD and SXRPD patterns recorded at 200 and 10 K was used to attempt a reliable refinement of nuclear and magnetic structures of the HT and LT phases of $\varepsilon$-$Fe_2O_3$, respectively. Besides $\varepsilon$-$Fe_2O_3$, the nuclear and magnetic structures of $\alpha$-$Fe_2O_3$, representing less than 10 wt.% of the sample were also considered in the refinements. The comparison between the experimental and calculated data obtained with the joint Rietveld refinement of the SXRPD and NPD patterns at 200 K and 10 K are presented in Figs. 3 and 4, respectively. The refined atomic coordinates for the 10 nonequivalent atomic sites at 200 and 10 K, together with the standard reliability factors are given in Table I. As it is found for the isostructural phases $GaFeO_3$, $AlFeO_3$ and $\kappa$-$Al_2O_3$ which have been previously refined,[16,12,17] the cation coordination polyhedra are two distorted[18] and one regular octahedra and one tetrahedron. Therefore, the different $Fe^{3+}$ sites are hereafter referred as $Fe_{DO1}$, $Fe_{DO2}$, $Fe_{RO}$, and $Fe_T$ (where DO stands for distorted octahedral, RO for regular octahedral and T for tetrahedral coordinations). In figure 5, the coordination octahedra of $Fe_{DO1}$ and $Fe_{DO2}$ are respectively represented in black and dark grey whereas the $Fe_{RO}$ and $Fe_T$ environments are represented in light grey. As expected, the changes between the asymmetric units of the HT and LT phases are in general small, although the $z$ coordinate of O2 significantly increases on cooling from 200 to 10 K (see Table I). As can be seen from Fig. 5 (black spheres), the 0.3 Å displacement of O2 along $c$ is responsible for the increased tilting of the $Fe_{DO1}$ and $Fe_T$ coordination polyhedra in the LT structure. The bonding distances and angles calculated from cell parameters and asymmetric units reported in Table I (see Tables II and III) reveal other features that can be relevant to understand the HT→LT magnetic transition. On the one hand, it is found that on cooling from 200 to 10 K there is an increase in the bonding distances of both $Fe_T$ with O4, O6 and $Fe_{DO2}$, $Fe_{RO}$ with O2, the latter showing an average expansion of about 1 %. In the case of the tetrahedral site, the elongation of the apical $Fe_T$-O4 distance shows a 2.2 % variation. On the other hand, the distortions of the $Fe_T$ and $Fe_{DO1}$ polyhedra, which are the more affected by the HT→LT transition, increase from $2.4 \cdot 10^{-4}$ to $6.8 \cdot 10^{-4}$ and from $71.3 \cdot 10^{-4}$ to $84.3 \cdot 10^{-4}$, respectively, the $Fe_{DO1}$-O2 distance decreasing by about 10 %.

Regarding the magnetic structures, the model that yielded the best fits for the HT phase is one in which the magnetic moments, *m*, are along the *a* axis with $m_{HT}(Fe_{DO1})$= -3.9$\mu_B$, $m_{HT}(Fe_{DO2})$= 3.9$\mu_B$, $m_{HT}(Fe_T)$= -2.4$\mu_B$, and $m_{HT}(Fe_{RO})$= 3.7$\mu_B$. The refinement was performed using the constraint $m_{HT}(Fe_{DO1}) = -m_{HT}(Fe_{DO2})$ suggested by Mössbauer spectroscopy measurements (see section B below). Thus, the ε-$Fe_2O_3$ HT magnetic structure presents the *m'm2'* point group and is represented in Fig. 6. The antiferromagnetic coupling between the different magnetic moments yields a net magnetization of 0.3$\mu_B$ per $Fe^{3+}$, in agreement with the $M_s$ values of ε-$Fe_2O_3$ at 200 K.[9] It is interesting to note that the NPD data can be also simulated allowing small *y* and *z* components for the $Fe_T$ and $Fe_{RO}$ momenta (*i.e.*, a small canting) although this did not result in any improvement of the fit. However, the fact that the NPD data does not exclude small canting of the $Fe_T$ and $Fe_{RO}$ spins would be in agreement with the thermal fluctuations in these sites, recently reported by Tronc *et al.*[19] For the LT phase, different helimagnetic and sine modulated structures have been tested to account for the magnetic structure associated with *k*. The best fit has been obtained for a sine modulated structure with a periodicity of about 10 crystalline unit cells, and all the magnetic moments laying in the *xy* plane. As in the HT structure, $m_{LT}(Fe_{DO1})$ and $m_{LT}(Fe_{DO2})$ were constrained to be antiparallel and directed along the *x* direction, whereas $m_{LT}(Fe_T)$ and $m_{LT}(Fe_{RO})$ are mainly oriented in opposite directions along *x* and present small *y* components. However, for the LT structure the symmetry analysis imposes that all the magnetic atoms are split into two orbits and the fact that the structure is sine-modulated introduces magnetic phases. This results in an increased number of refinable magnetic parameters that increase the difficulty at finding an appropriate model for the magnetic structure from the small number of measurable magnetic reflections. All the magnetic moments obtained in our best fit of the NPD pattern at LT using a simple sine wave modulated structure increase with respect to those of the HT structure. The values of some of them are, however, clearly non-physical, exceeding the $5\mu_B$ expected for a $Fe^{3+}$ cation.

**B-Mössbauer Spectroscopy**

Mössbauer spectroscopy gives additional information that can be complementary to the NPD data. In Fig. 7 it can be observed how the Mössbauer spectra of $\epsilon$-$Fe_2O_3$ change on cooling through the HT→LT transition, especially due to the evolution of the $Fe_T$ subspectrum (dotted line). According to the crystalline structure of $\epsilon$-$Fe_2O_3$, the spectra were fitted assuming each of the four Fe sites contributing to the spectra with a sextet. In particular, the spectra could be fitted by constraining the hyperfine fields, $B_{hf}$, and the widths of the $Fe_{DO1}$ and $Fe_{DO2}$ contributions to be equal, in accordance with the NPD results. Fig. 8a shows the temperature dependence of the hyperfine parameters for the different Fe sites in the 10 K to 200 K range. Between 150 and 80 K most of the hyperfine parameters of the different Fe sites deviate from the thermal dependence displayed at higher temperatures. These anomalies are particularly important for $Fe_T$, which displays a 20 % increase in $B_{hf}$ between 140 and 100 K, and a shift in both the isomer shift, $\delta$, and the quadrupole splitting, $\Delta$, in the same temperature range. Below 150 K, $B_{hf}$ decreases by ~ 4 % for $Fe_{RO}$ between 150 and 110 K and by ~3.5% for $Fe_{DO1}$ and Fe$_{DO2}$ between 150 and 130 K. Note that for Fe$_{DO1}$, $\Delta$ changes from negative to positive values in this interval of temperatures and that below 130 K, $\delta$ decreases with temperature for this particular site, contrasting with the usual temperature dependence of the isomer shift. The latter results suggest that on cooling below 150 K some structural changes would affect the Fe$_{DO1}$ coordination octahedron which, subsequently, would induce changes in the coordination of the Fe$_T$ site. Indeed, this interpretation is compatible with the results obtained from the diffraction experiments which revealed that the HT→LT transition results in an increased tilting and distortion of the Fe$_{DO1}$ and Fe$_T$ coordination polyhedron. It is worth pointing out that large changes in $\Delta$ can be also due to spin reorientations. Although the NPD results seem to indicate that the HT→LT has little effect in the direction of the Fe$_{DO1}$ a magnetic moment, LT magnetic structure would be needed to ascribe the sign change in $\Delta$ for this site exclusively to structural changes.

The structural transformations occurring below 150 K might be responsible for the magnetic softening of the system that is reported in Fig. 1. Interestingly, the changes observed in $\Delta$ for both the Fe$_{DO1}$ and Fe$_T$ sites can be related to modifications in the spin-orbit coupling of the Fe cations which are responsible for the magnetic anisotropy of a magnetic system. In contrast, in the case of the Fe$_T$ site, the substantial elongation of the apical Fe$_T$-O4 occurring along the HT→LT transformation could be responsible for the large increase of $B_{hf}$ observed at this site which, in turn, indicates an increment of the Fe$_T$ magnetic moment,[20] also deduced from the analysis of the NPD data at 10 K. This can explain the $M_s$ diminution observed below 150 K (see Fig. 1 inset) since the Fe$_T$ and Fe$_{RO}$ are antiferromagnetically coupled and the increase of $B_{hf}$ at the Fe$_T$ site is larger than the $B_{hf}$ increase experienced by Fe$_{RO}$. Regarding the hyperfine fields, it is also interesting to note that the ratio between the $B_{hf}$ values for different Fe sites at 200

K [*i.e.* $B_{hf}$ (Fe$_{DO1}$)/$B_{hf}$ (Fe$_T$)=1.63, $B_{hf}$ (Fe$_{RO}$)/$B_{hf}$ (Fe$_T$)=1.56] is in good agreement with the corresponding ratio of magnetic moments obtained from the analysis of the NPD data. Assuming that the proportionality between $B_{hf}$ and $m$ still holds for the LT phase and combining the $m_{HT}$ values obtained from the NPD pattern at 200 K with the $B_{hf}$ at 10 K, the magnetic moments for the LT phase can be estimated to be $m_{LT}$(Fe$_{DO1}$)= -4.2$\mu_B$, $m_{LT}$(Fe$_{DO2}$)= 4.2$\mu_B$, $m_{LT}$(Fe$_T$)= -3.7$\mu_B$ and $m_{LT}$(Fe$_{RO}$)= 4.1$\mu_B$, in clear disagreement with the unrealistic values obtained from our best fit model. In this regard, the temperature dependence of the Mössbauer subspectra linewidth (see Fig. 8b) can shed some light on the deficiencies of our preliminary model for the LT magnetic structure of ε-Fe$_2$O$_3$. On cooling below 150 K a sextet linewidth broadening is observed for all Fe sites but specially for Fe$_T$ which is due to the appearance of some disorder at the different crystallographic sites and indicates that the transformation is rather broad and that several phases can coexist in the 150-92 K interval. However, as $T$ is further lowered decreased the linewidth values decrease progressively. Interestingly, at about 10 K the linewidth values for all the sites are only slightly larger than those at 200 K, implying that the Mössbauer components for all the sites are quite narrow and, consequently, that the corresponding magnetic moments throughout the structure should be almost constant in modulus. However, this is not what one would expect if the LT magnetic structure of ε-Fe$_2$O$_3$ is supposed to be sine-modulated with a periodicity of about 10 unit cells. In this case, since for the *Pna2$_1$* space group there is only one Wyckoff position of fourfold multiplicity, an average of 40 atoms for every Fe site with magnetic moments ranging from zero to a maximum value are involved in one period of the amplitude-modulated structure which would yield a broad hyperfine field distribution. Since the helical or spiral structures with constant $m$ modulus that could account for the narrow Mössbauer lines of the LT phase are not consistent with the NPD data, the only magnetic structure that can make compatible both the Mössbauer and the diffraction data is a square-wave modulated structure, *i.e.* the superposition of a series of sine-modulated structures having the harmonics of **k** as propagation vectors. In fact, the exceedingly large magnetic moments obtained when a single sine-wave was considered might be reduced by taking into account the amplitudes of the different Fourier components. Thus, in order to determine the exact magnetic structure it would be necessary to identify the higher order reflections and determine the new propagation vectors but this cannot be carried out due to the peak overlapping and the low intensity of the harmonic reflections. However, on the grounds of the above discussion one may consider the structure represented in Fig. 9 to be a simplified schematic model of the LT magnetic cell.

### C. Magnetization and heat capacity measurements

The results presented in section B indicate that the HT→LT transformation in $\varepsilon$-Fe$_2$O$_3$ takes place in a broad temperature range and suggest that it consists in a succession of magnetic and structural changes. In this section we present magnetic and heat capacity measurements that give some details about the nature of the transition. Fig. 10a shows the temperature dependence of the $\varepsilon$-Fe$_2$O$_3$ zero-field-cooling (ZFC) magnetization under 100 and 1000 Oe applied fields, $M_{100}^{ZFC}(T)$ and $M_{1000}^{ZFC}(T)$, respectively, together with the field-cooling (FC) measurement in 1000 Oe, $M_{1000}^{FC}(T)$. Upon heating from 10 K, $M_{100}^{ZFC}(T)$ presents an important increase between 85 and 150 K which takes place in three stages, if one considers the changes of the $M_{100}^{ZFC}(T)$ slope, in three temperature ranges $\Delta T_1$ (85 K<$T$<95 K), $\Delta T_2$ (95 K<$T$<110 K) and $\Delta T_3$ (110 K<$T$< 147 K), the slope being much steeper in $\Delta T_1$ and $\Delta T_2$ than in $\Delta T_3$. When the ZFC magnetization is measured in a field 10 times larger (see $M_{1000}^{ZFC}(T)$ curve), the larger magnetization step starts at a lower temperature, is more gradual and it is no longer possible to clearly distinguish between the first two stages. In contrast, no shift is observed for the magnetization stage in $\Delta T_3$, which now presents a somewhat larger slope. Moreover, the FC measurement (see $M_{1000}^{FC}(T)$ curve) shows that the magnetic behaviour below 110 K is irreversible even for small fields such as 1000 Oe, in contrast to the reversibility observed above this temperature. These measurements suggest that on heating, the transition from the LT to the HT magnetic structure of ε-$Fe_2O_3$ takes place in three consecutive stages occurring in $\Delta T_1$, $\Delta T_2$ and $\Delta T_3$, hereafter referred as S1, S2 and S3, respectively. The measurements of Fig. 10a evidence a distinct magnetic character of S1 and S2 as compared to S3 which is also confirmed by the temperature dependence of the magnetic AC susceptibility, $\chi$. Namely, in Fig. 10b, the temperature dependence of both the in-phase, $\chi'(T)$, and out of phase, $\chi''(T)$, components of AC susceptibility present sharp peaks at about 91 and 101 K signalling the S1 and S2 transformations, respectively. Indeed, the susceptibility peaks coincide with the maxima in the temperature derivative of $M_{100}^{ZFC}(T)$ (solid line in Fig. 10b), *i.e.* with the maximum slopes of $M_{100}^{ZFC}(T)$ in $\Delta T_1$ and $\Delta T_2$. On the other hand, for $\Delta T_3$, only a hardly observable shoulder occurs in $\chi'(T)$ while $\chi''(T)$ is zero. The magnetic measurements presented in Fig. 10 are in agreement with the occurrence of magnetic and structural changes along $\Delta T_3$ that can be inferred from the anomalies in the Mössbauer hyperfine parameters in this temperature interval. Thus, S3 is to be associated to the latter transformation which in turn, induces S2 and S1 at lower temperatures, which are predominantly related to a magnetic transition displaying a field-dependent behavior.

The distinct magnetic nature of S3 on the one hand and S2+S1 on the other is also revealed in Fig. 11 by the temperature dependence of the NPD patterns. On cooling below 200 K, the intensity of the (120) magnetic reflection slightly decreases, this diminution becoming much more pronounced below 150 K, coinciding with the gradual appearance of the satellites that characterize an incommensurate order (S3). Below 110 K, the intensity of the satellites suddenly increases (S2+S1) resulting in the LT magnetic structure of $\varepsilon$-$Fe_2O_3$ which is fully developed at about 80 K. Heat capacity measurements provide an additional evidence of the different characteristics of S3 with respect of S2 and S1. The temperature dependence of $C_p/T$ represented in Fig. 12 shows an anomaly around 130 K (*i.e.* in S3) and two distinct regimes above and below this temperature. Since $\frac{C_p}{T} = -\frac{\partial^2 G}{\partial T^2}$, where $G$ is the free energy, the anomaly in $C_p/T$ suggests that S3 is related with a second order magnetic and structural transition which is supported by the absence of thermal hysteresis both in the magnetic (Fig. 10) and dielectric[10] measurements.

## IV. CONCLUSION

In conclusion, the HT magnetic structure of $\varepsilon$–$Fe_2O_3$ is that of a collinear ferrimagnetic material with the $Fe^{3+}$ magnetic moment antiferromagnetically coupled along *a*. The $Fe^{3+}$ magnetic moments in the $Fe_{DO1}$ and $Fe_{DO2}$ distorted octahedral positions mutually cancel and the net magnetization of this phase results from the uncompensated magnetic moment of the atoms in tetrahedral ($Fe_T$) and regular octahedral ($Fe_{RO}$) positions. Between 150 and 80 K, $\varepsilon$–$Fe_2O_3$ undergoes magnetic and structural phase transformations which bring about a gradual decrease of the magnetic anisotropy. This transition takes place at least in three stages. Between 150 and 110 K there are evidences of a second order structural transition presumably involving the changes in the coordination of the $Fe_{DO1}$ and $Fe_T$ sites, occurring simultaneously with the emergence of an incommensurate magnetic order. The magnetic structure undergoes several transformations as the temperature is further decreased but no additional changes are observed below 80 K. The combination of powder diffraction and Mössbauer spectroscopy measurements indicate that $\varepsilon-Fe_2O_3$ presents a square-wave incommensurate magnetic structure at LT.

**ACKNOWLEDGEMENTS**

Financial support from Ministerio de Educación y Ciencia, Projects MAT2003-01052 and MAT2004-01679, Generalitat de Catalunya, Projects 2005SGR00452 and 2005SGR00401, are gratefully acknowledged. C. F. acknowledges financial support from MEC (Spain). W. A. A. M. acknowledges support from CNPq (Brazil). We acknowledge the ESRF and the ILL for the provision of X-ray and neutron beamtime. We also thank F. Fauth for his assistance during XRD data collection.

[18] The distortion of a coordination polyhedron of a cation is defined as $\frac{1}{n}\sum_{i=1}^{n}\left[\frac{d_i - <d>}{<d>}\right]^2$, where $d_i$ is de distance to a given neighbour, $<d>$ the average distance to the first neighbours, and $n$ is the coordination number. At 200 K, the distortions of the $Fe_{DO-1}$, $Fe_{DO-2}$, $Fe_{RO}$, and $Fe_T$ have been found to be $71.3 \cdot 10^{-4}$, $58.4 \cdot 10^{-4}$, $2.4 \cdot 10^{-4}$ and $4.6 \cdot 10^{-4}$, respectively.

**FIGURE CAPTIONS**

**FIG. 1:** Temperature dependence of the coercive field, $H_C$, the squareness ratio, $M_R/M_S$ and the saturation magnetization, $M_S$, in $\varepsilon$-Fe$_2$O$_3$ (taken from Ref. 9).

**FIG. 2:** NPD patterns corresponding to the HT (200, 160 K) and LT phase (80, 10 K) of $\varepsilon$-Fe$_2$O$_3$. The up arrows indicate the satellites emerging in the LT phase.

**FIG. 3:** Experimental (■), calculated (solid line) and difference plot (lower line) for the joint refinement of SXRPD and NPD patterns of $\varepsilon$-Fe$_2$O$_3$ collected at 200 K (upper and lower panel respectively). Reflection positions of the different phases are indicated by vertical bars in the following descending order: $\varepsilon$-Fe$_2$O$_3$, $\alpha$-Fe$_2$O$_3$ for the upper panel and $\varepsilon$-Fe$_2$O$_3$, $\varepsilon$-Fe$_2$O$_3$ magnetic, $\alpha$-Fe$_2$O$_3$, $\alpha$-Fe$_2$O$_3$ magnetic for the lower panel.

**FIG. 4:** Experimental (■), calculated (solid line) and difference plot (lower line) for the joint refinement of SXRPD and NPD patterns of $\varepsilon$-Fe$_2$O$_3$ collected at 10 K (upper and lower panel respectively). Reflection positions of the different phases are indicated by vertical bars in the following descending order: $\varepsilon$-Fe$_2$O$_3$, $\alpha$-Fe$_2$O$_3$ for the upper panel and $\varepsilon$-Fe$_2$O$_3$, $\varepsilon$-Fe$_2$O$_3$ magnetic, $\alpha$-Fe$_2$O$_3$, $\alpha$-Fe$_2$O$_3$ magnetic for the lower panel.

**FIG. 5:** Perspective view of the ε-Fe$_2$O$_3$ HT and LT structures along the [100] direction (upper and lower panel, respectively). The coordination octahedra of Fe$_{DO1}$, Fe$_{DO2}$ and Fe$_{RO}$ are black, grey and white, respectively and the oxygen atoms are represented by small spheres. The main structural changes concern the orientations of the coordination polyhedra of Fe$_{DO1}$ (black octahedron) and Fe$_T$ (light grey tetrahedron) due to de displacement of O2 (in black) along *c*.

**FIG. 6:** Magnetic structure of the ε-Fe$_2$O$_3$ HT phase where the Fe$_{DO1}$, Fe$_{DO2}$, Fe$_{RO}$ and Fe$_T$ are distinguished by the 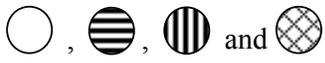 hachure patterns, respectively. The magnetic moments of the Fe$_{DO1}$, Fe$_{DO2}$, Fe$_{RO}$ and Fe$_T$ sites are schematically represented in the figure by black, grey, light grey and dotted arrows, respectively.

**FIG. 7:** Mössbauer spectra of ε-Fe$_2$O$_3$ at 200, 120 and 92 K. The subspectra correspond to the different Fe sites in the ε-Fe$_2$O$_3$ structure and are schematically presented at the top of the figure.

**FIG. 8:** Temperature dependence of the hyperfine field (B$_{hf}$), sextet linewidth (width), isomer shift (δ) and quadrupolar splitting (Δ) of the different Fe sites of ε-Fe$_2$O$_3$ obtained from Mössbauer spectra.

**FIG. 9:** Simplified schematic representation of the square-wave magnetic structure of the LT phase of ε-Fe$_2$O$_3$. The magnetic moments of the Fe$_{DO1}$, Fe$_{DO2}$, Fe$_{RO}$ and Fe$_T$ sites are schematically represented by black, grey, light grey and dotted arrows, respectively.

**FIG. 10:** (a): Zero field cooling (ZFC) and field cooling (FC) measurements of the temperature dependence of the magnetization in 100 and 1000 Oe applied fields. (b): Temperature dependence of the in-phase (▲) and out-of-phase (△) components of ε-Fe$_2$O$_3$ ac-susceptibility measured at 1000 Hz and 10 Oe field amplitude. The solid line is the temperature derivative of the $M_{100}^{ZFC}(T)$ curve presented of panel (a).

**FIG. 11:** Temperature dependence of the (120) magnetic reflection and its satellites of ε-Fe$_2$O$_3$.

**FIG. 12:** Temperarture dependence of the ε-Fe$_2$O$_3$ heat capacity over the temperature.

**TABLES**

**Table I**

| $T = 200$ | | | | $T = 10$ K | | | |
|---|---|---|---|---|---|---|---|
| | $a$=5.0885(5) | $b$=8.7802(14) | $c$=9.4709(13) | | $a$=5.085(1) | $b$=8.774(2) | $c$=9.468(2) |
| Atom | x/a | y/b | z/c | Atom | x/a | y/b | z/c |
| O1 | 0.978(2) | 0.3282(15) | 0.4314(11) | O1 | 0.978(4) | 0.331(3) | 0.4288(17) |
| O2 | 0.515(2) | 0.4907(17) | 0.4187(16) | O2 | 0.512(3) | 0.4871(20) | 0.4489(14) |
| O3 | 0.650(3) | 0.9979(13) | 0.1883(9) | O3 | 0.646(4) | 0.9943(20) | 0.1871(14) |
| O4 | 0.160(3) | 0.1637(15) | 0.1956(7) | O4 | 0.159(4) | 0.162(2) | 0.2002(11) |
| O5 | 0.841(3) | 0.1680(15) | 0.6669(7) | O5 | 0.858(3) | 0.157(3) | 0.6685(12) |
| O6 | 0.527(2) | 0.1637(19) | 0.9362(9) | O6 | 0.523(4) | 0.161(3) | 0.9257(15) |
| $Fe_{DO1}$ | 0.1928(11) | 0.1506(6) | 0.5807(3) | $Fe_{DO1}$ | 0.1931(17) | 0.1514(10) | 0.5820(4) |
| $Fe_{DO2}$ | 0.6826(6) | 0.0291(3) | 0.7897(5) | $Fe_{DO2}$ | 0.6867(10) | 0.0283(4) | 0.7938(8) |
| $Fe_T$ | 0.1858(10) | 0.1519(6) | 0.0000 | $Fe_T$ | 0.1852(15) | 0.1526(9) | 0.0000 |
| $Fe_{RO}$ | 0.8104(7) | 0.1580(4) | 0.3071(3) | $Fe_{RO}$ | 0.8098(10) | 0.1592(6) | 0.3079(5) |
| SXRPD: $R_B$=2.26, $R_{wp}$= 3.59, $R_{exp}$= 1.94 $\chi^2$=3.43  NPD: $R_B$=1.41, $R_{Mag}$= 1.89, $R_{wp}$= 4.00, $R_{exp}$= 1.33 | | | | SXRPD: $R_B$=2.98, $R_{wp}$= 7.33, $R_{exp}$= 3.66, $\chi^2$=4.015  NPD: $R_B$=3.85, $R_{Mag}$= 6.33, $R_{wp}$= 7.60, $R_{exp}$= 1.40 | | | |

Cell parameters and assymetric units of $\varepsilon$-$Fe_2O_3$ at 200 and 10 K, together with the reliability parameters of the multipattern Rietveld refinement expressed in % (except for $\chi^2$).

M. Gich *et al*.

**Table II**

| O1 | Fe$_{DO1}$ | Fe$_{DO1}$[b] | Fe$_{DO2}$[a] | Fe$_{RO}$ | Fe$_{RO}$[b] |
|---|---|---|---|---|---|
| Fe$_{DO1}$ | **2.372(13)** *2.41(2)* | | | | |
| Fe$_{DO1}$[b] | 88.5(5)  *86.8(9)* | **2.035(11)** *2.056(20)* | | | |
| Fe$_{DO2}$[a] | 171.5(5) *171.9(9)* | 94.6(5) *94.5(9)* | **2.362(13)** *2.31(2)* | | |
| Fe$_{RO}$ | 93.1(5) *92.5(9)* | 99.6(5) *98.8(9)* | 94.2(5) *95.1(9)* | **2.085(12)** *2.08(2)* | |
| Fe$_{RO}$[b] | 90.1(5) *89.4(8)* | 167.2(5) *167.2(9)* | 85.2(5) *87.6(8)* | 93.2(5) *93.5(9)* | **2.064(11)** *2.041(20)* |

| O2 | Fe$_{DO1}$[b] | Fe$_{DO2}$[a] | Fe$_T$[a] | Fe$_{RO}$[b] |
|---|---|---|---|---|
| Fe$_{DO1}$[b] | **2.171(15)** *1.976(17)* | | | |
| Fe$_{DO2}$[a] | 102.0(6) *102.1(8)* | **1.994(13)** *2.149(16)* | | |
| Fe$_T$[a] | 111.2(7) *125.0(10)* | 122.4(6) *116.0(8)* | **1.908(15)** *1.833(18)* | |
| Fe$_{RO}$[b] | 102.7(6) *104.8(7)* | 101.0(6) *91.0(7)* | 115.1(7) *112.2(8)* | **1.976(14)** *2.121(16)* |

| O3 | Fe$_{DO1}$[c] | Fe$_{DO2}$[c] | Fe$_{RO}$ |
|---|---|---|---|
| Fe$_{DO1}$[c] | **1.838(12)** *1.817(19)* | | |
| Fe$_{DO2}$[c] | 124.2(7) *126.6(10)* | **1.960(14)** *1.976(20)* | |
| Fe$_{RO}$ | 129.8(6) *128.6(9)* | 99.4(6)  *97.8(8)* | **1.977(12)** *2.026(18)* |

| O4 | Fe$_{DO2}$[c] | Fe$_T$ | Fe$_{RO}$ | Fe$_{RO}$[b] |
|---|---|---|---|---|
| Fe$_{DO2}$[c] | **2.074(13)** *2.047(18)* | | | |
| Fe$_T$ | 110.8(5) *111.7(7)* | **1.860(7)** *1.902(11)* | | |
| Fe$_{RO}$ | 95.3(6) *96.1(8)* | 124.6(5) *123.8(7)* | **2.069(14)** *2.048(19)* | |
| Fe$_{RO}$[b] | 105.0(5) *105.6(7)* | 122.2(5) *120.6(7)* | 94.4(6) *95.0(8)* | **2.038(13)** *2.022(17)* |

| O5 | Fe$_{DO1}$ | Fe$_{DO1}$[b] | Fe$_{DO2}$ |
|---|---|---|---|
| Fe$_{DO1}$ | **1.973(15)** *1.891(17)* | | |
| Fe$_{DO1}$[b] | 104.0(7) *102.5(10)* | **1.942(14)** *2.05(2)* | |
| Fe$_{DO2}$ | 126.8(7) *133.2(10)* | 129.0(6) *124.2(10)* | **1.868(12)** *1.855(20)* |

| O6 | Fe$_{DO2}$ | Fe$_T$ | Fe$_T$[b] |
|---|---|---|---|
| Fe$_{DO2}$ | **1.987(13)** *1.90(2)* | | |
| Fe$_T$ | 124.8(6) *129.0(10)* | **1.841(11)** *1.86(2)* | |
| Fe$_T$[b] | 123.9(7) *124.2(11)* | 110.1(7) *106.6(11)* | **1.908(16)** *1.96(3)* |

a: (½-x, y-½, z-½); b: (½-x, ½-y, z)

c: (-x, -y, z+½)

Coupling angles and distances (given in Å) between the different Fe sites at 200 and 10 K (in italics)

M. Gich *et al.*

**Table III**

|  | 200 K | 10 K |  | 200 K | 10 K |
|---|---|---|---|---|---|
| $O2^a$-$Fe_T$-O4 | 113.9(10) | 105.0(12) | O1-$Fe_{DO1}$-$O1^b$ | 81.6(7) | 81.1(13) |
| $O2^a$-$Fe_T$-O6 | 114.5(10) | 116.1(17) | O1-$Fe_{DO1}$-$O2^b$ | 76.2(8) | 78.9(13) |
| $O2^a$-$Fe_T$-$O6^b$ | 106.0(13) | 109.5(18) | O1-$Fe_{DO1}$-$O3^c$ | 175.9(10) | 175.8(17) |
| O4-$Fe_T$-O6 | 113.0(8) | 116.1(13) | O1-$Fe_{DO1}$-O5 | 77.2(7) | 80.5(12) |
| O4-$Fe_T$-$O6^b$ | 103.8(9) | 107.0(13) | O1-$Fe_{DO1}$-$O5^b$ | 83.7(7) | 83.7(12) |
| O6-$Fe_T$-$O6^b$ | 104.3(10) | 102.7(18) | $O1^b$-$Fe_{DO1}$-$O2^b$ | 81.9(8) | 85.6(13) |
|  |  |  | $O1^b$-$Fe_{DO1}$-$O3^c$ | 98.0(8) | 96.9(13) |
| O1-$Fe_{RO}$-$O1^b$ | 88.3(7) | 90.0(15) | $O1^b$-$Fe_{DO1}$-O5 | 158.0(11) | 159.8(15) |
| O1-$Fe_{RO}$-$O2^b$ | 87.5(9) | 83.8(14) | $O1^b$-$Fe_{DO1}$-$O5^b$ | 86.6(8) | 86.1(13) |
| O1-$Fe_{RO}$-O3 | 179.5(10) | 179.0(18) | $O2^b$-$Fe_{DO1}$-$O3^c$ | 99.7(10) | 97.3(12) |
| O1-$Fe_{RO}$-O4 | 85.4(8) | 84.8(12) | $O2^b$-$Fe_{DO1}$-O5 | 87.6(9) | 82.8(11) |
| O1-$Fe_{RO}$-$O4^b$ | 84.0(7) | 82.6(12) | $O2^b$-$Fe_{DO1}$-$O5^b$ | 158.1(12) | 161.6(17) |
| $O1^d$-$Fe_{RO}$-$O2^b$ | 99.5(9) | 94.3(13) | $O3^c$-$Fe_{DO1}$-O5 | 102.7(11) | 100.9(15) |
| $O1^d$-$Fe_{RO}$-O3 | 91.6(8) | 90.3(12) | $O3^c$-$Fe_{DO1}$-$O5^b$ | 100.4(10) | 100.0(17) |
| $O1^d$-$Fe_{RO}$-O4 | 173.7(11) | 174.7(17) | O5-$Fe_{DO1}$-$O5^b$ | 96.6(11) | 100.0(15) |
| $O1^d$-$Fe_{RO}$-$O4^b$ | 86.7(7) | 86.4(12) |  |  |  |
| $O2^c$-$Fe_{RO}$-O3 | 93.0(9) | 97.2(11) | $O1^a$-$Fe_{DO2}$-$O2^a$ | 78.0(7) | 75.7(11) |
| $O2^c$-$Fe_{RO}$-O4 | 80.6(9) | 84.2(10) | $O1^a$-$Fe_{DO2}$-$O3^c$ | 83.6(7) | 84.2(12) |
| $O2^c$-$Fe_{RO}$-$O4^b$ | 169.3(12) | 166.4(14) | $O1^a$-$Fe_{DO2}$-$O4^c$ | 76.6(7) | 76.5(11) |
| O3-$Fe_{RO}$-O4 | 94.7(10) | 94.9(14) | $O1^a$-$Fe_{DO2}$-O5 | 172.1(10) | 168.7(19) |
| O3-$Fe_{RO}$-$O4^b$ | 95.5(9) | 96.5(13) | $O1^a$-$Fe_{DO2}$-O6 | 84.8(9) | 86.4(17) |
| O4-$Fe_{RO}$-$O4^b$ | 92.3(10) | 93.9(14) | $O2^a$-$Fe_{DO2}$-$O3^c$ | 161.1(12) | 159.9(15) |
|  |  |  | $O2^a$-$Fe_{DO2}$-$O4^c$ | 80.1(8) | 83.5(10) |
|  |  |  | $O2^a$-$Fe_{DO2}$-O5 | 99.1(10) | 101.9(12) |
|  |  |  | $O2^a$-$Fe_{DO2}$-O6 | 88.9(9) | 88.0(13) |
|  |  |  | $O3^c$-$Fe_{DO2}$-$O4^c$ | 91.4(10) | 91.4(14) |
|  |  |  | $O3^c$-$Fe_{DO2}$-O5 | 98.4(11) | 97.8(15) |
|  |  |  | $O3^c$-$Fe_{DO2}$-O6 | 94.0(9) | 91.3(13) |
|  |  |  | $O4^c$-$Fe_{DO2}$-O5 | 95.7(10) | 92.3(16) |
|  |  |  | $O4^c$-$Fe_{DO2}$-O6 | 159.9(11) | 162.3(18) |
|  |  |  | O5-$Fe_{DO2}$-O6 | 102.6(8) | 104.7(13) |

a: (½-$x$, $y$-½, $z$-½); b: (½-$x$, ½-$y$, $z$)

c: (-$x$, -$y$, $z$+½)

Bonding angles in the coordination polyhedra for the four Fe sites of $\varepsilon$-$Fe_2O_3$ at 200 and 10 K.

M. Gich *et al*.

**Figure 1**

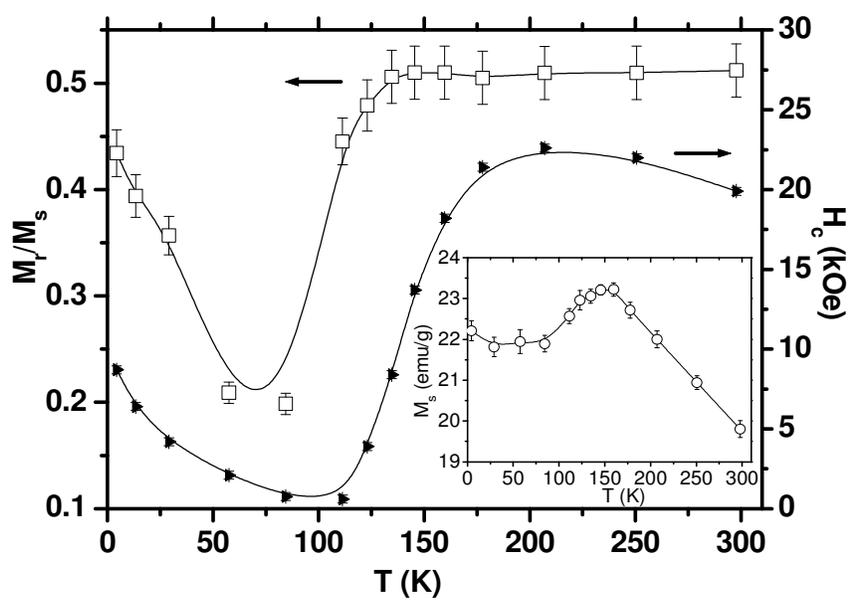

M. Gich *et al*.

**Figure 2**

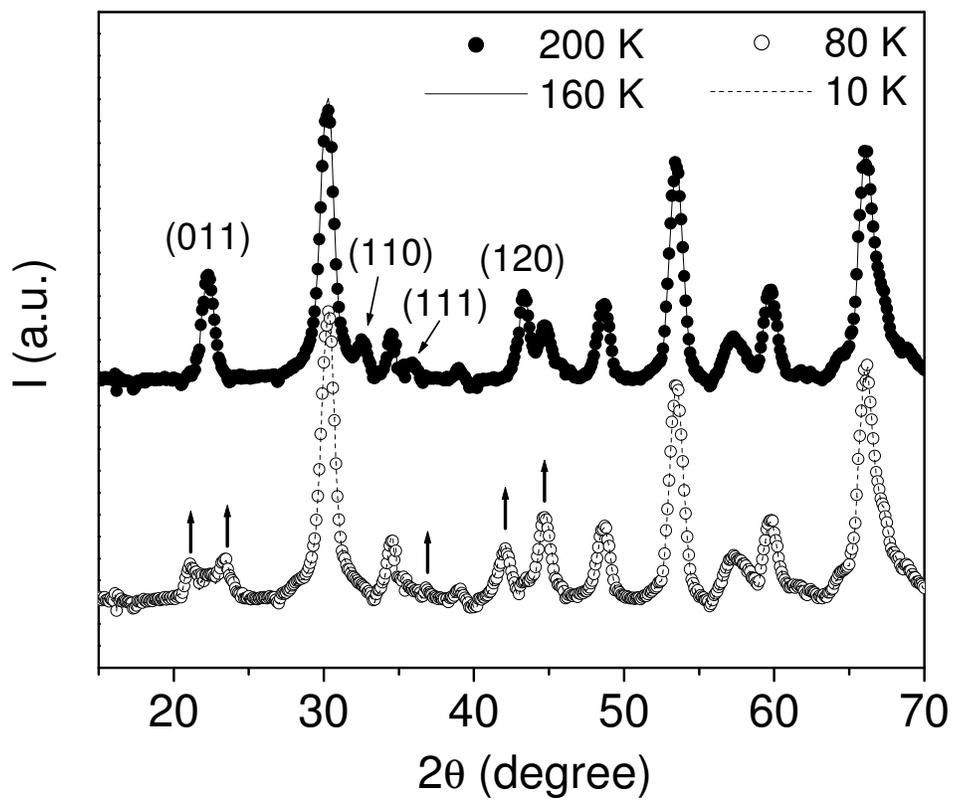

M. Gich *et al*.

**Figure 3:**

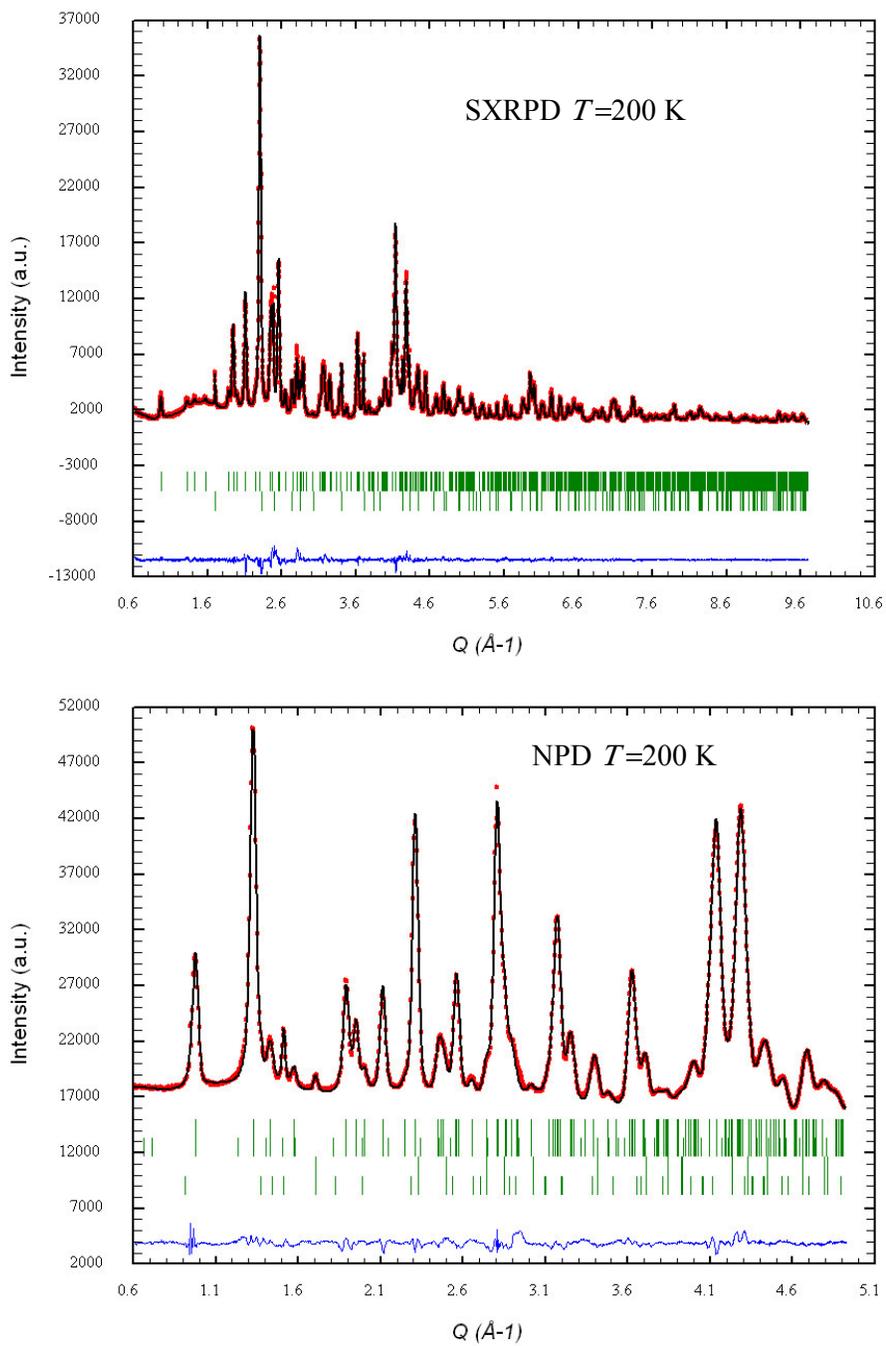

M. Gich *et al*.

**Figure 4:**

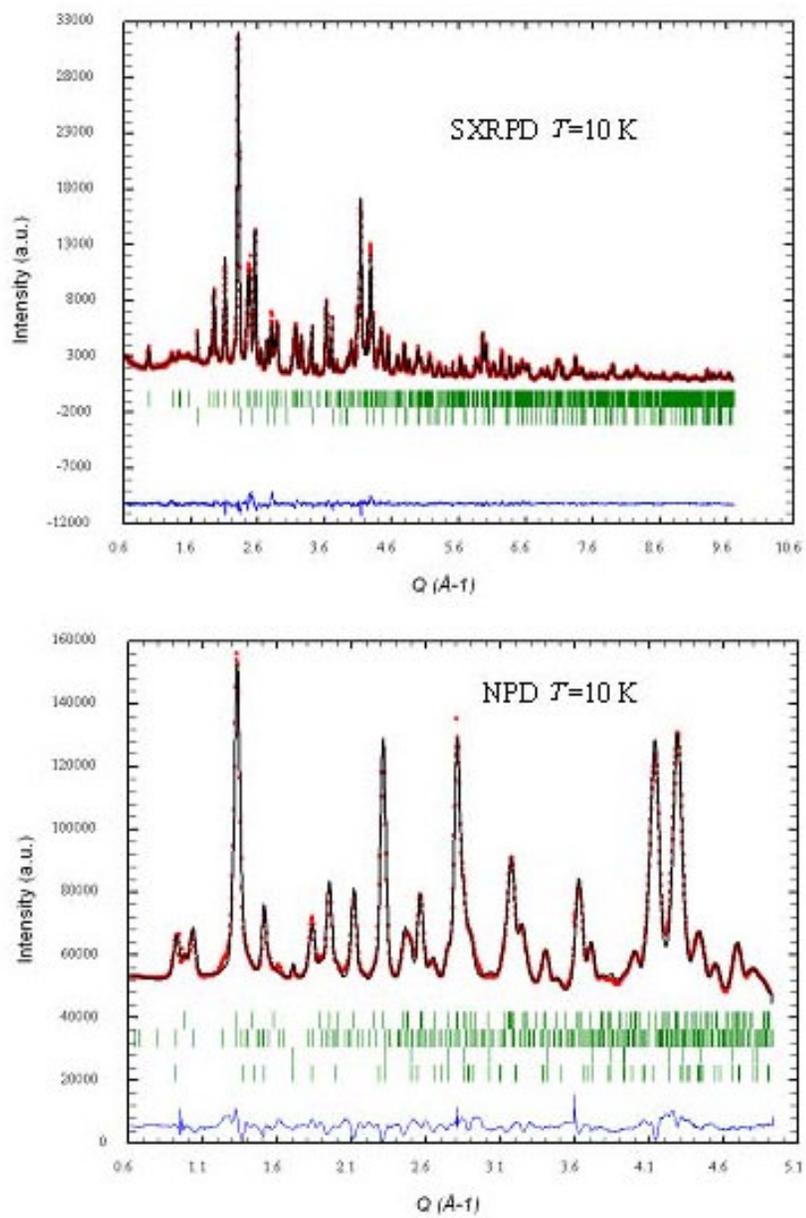

M. Gich *et al*.

**Figure 5:**

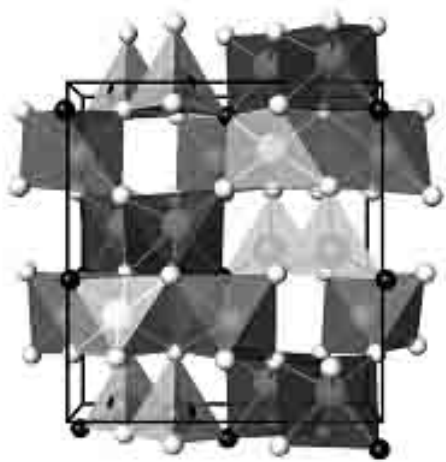

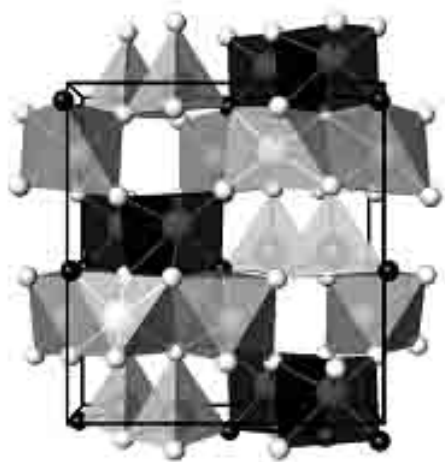

M. Gich *et al*.

**Figure 6:**

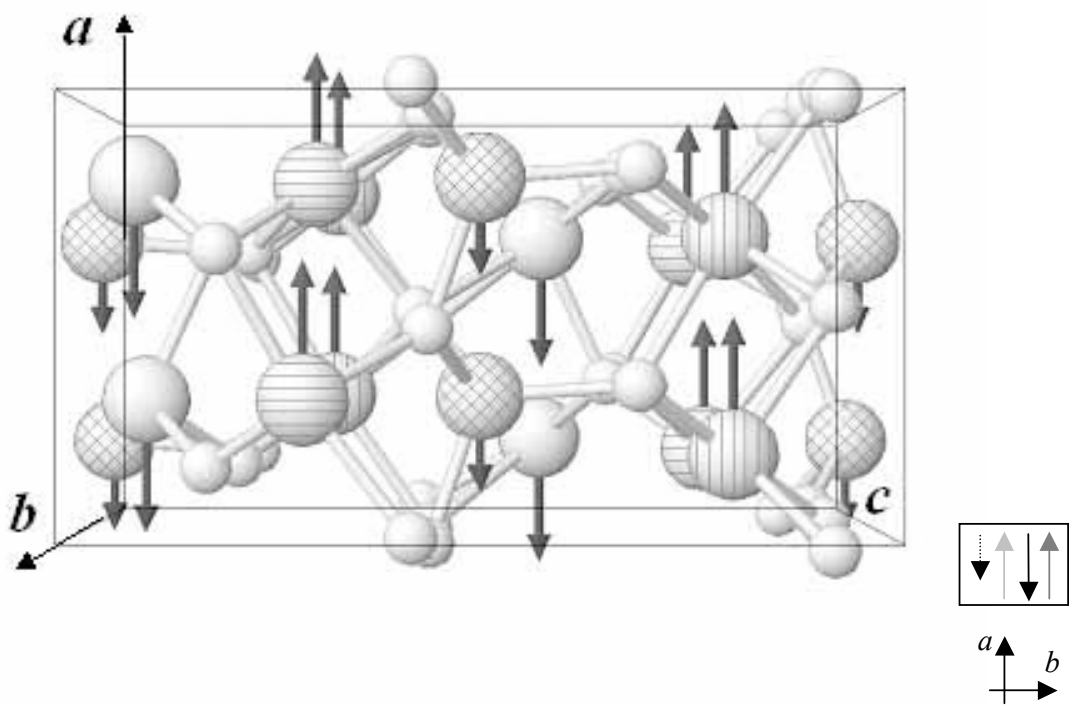

M. Gich *et al.*

**Figure 7**

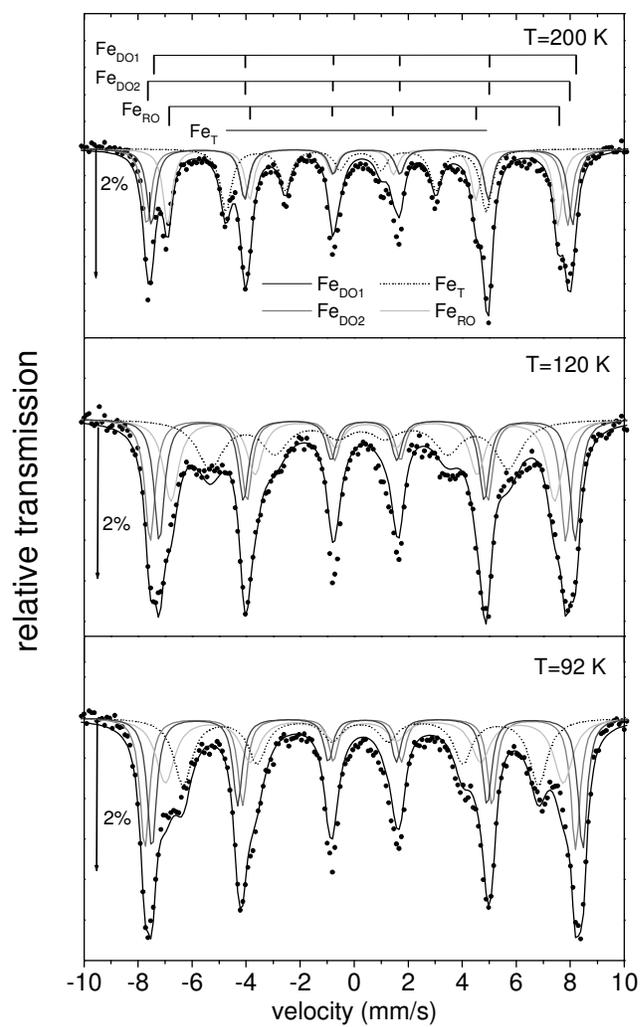

M. Gich *et al.*

**Figure 8**

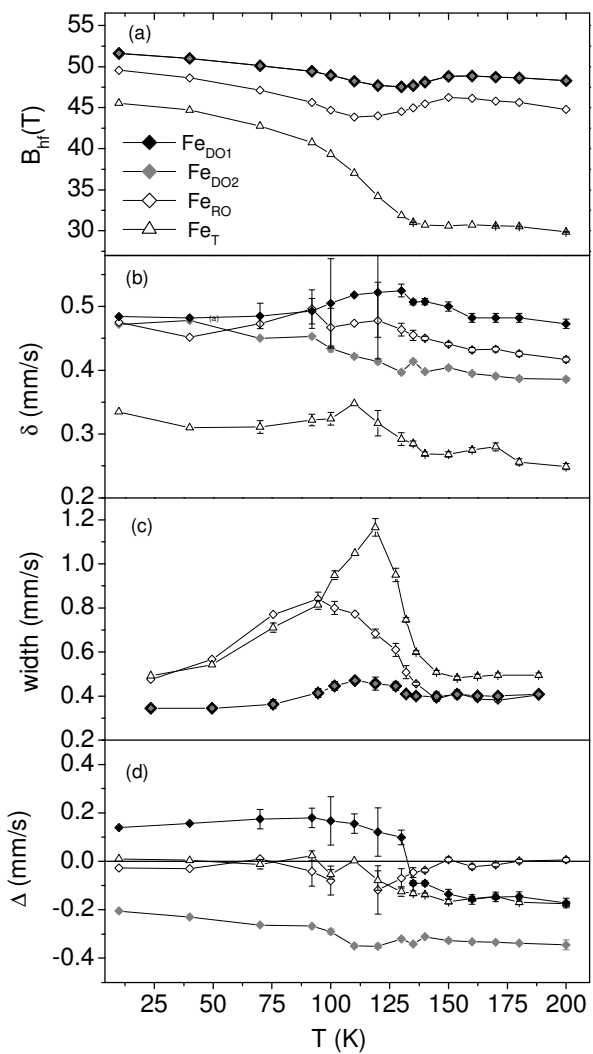

M. Gich *et al*.

**Figure 9**

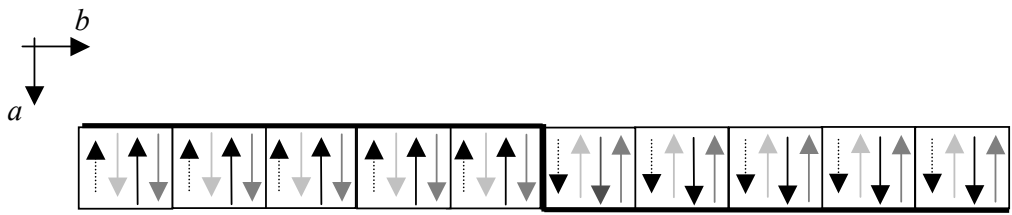

M. Gich et al.

**Figure 10**

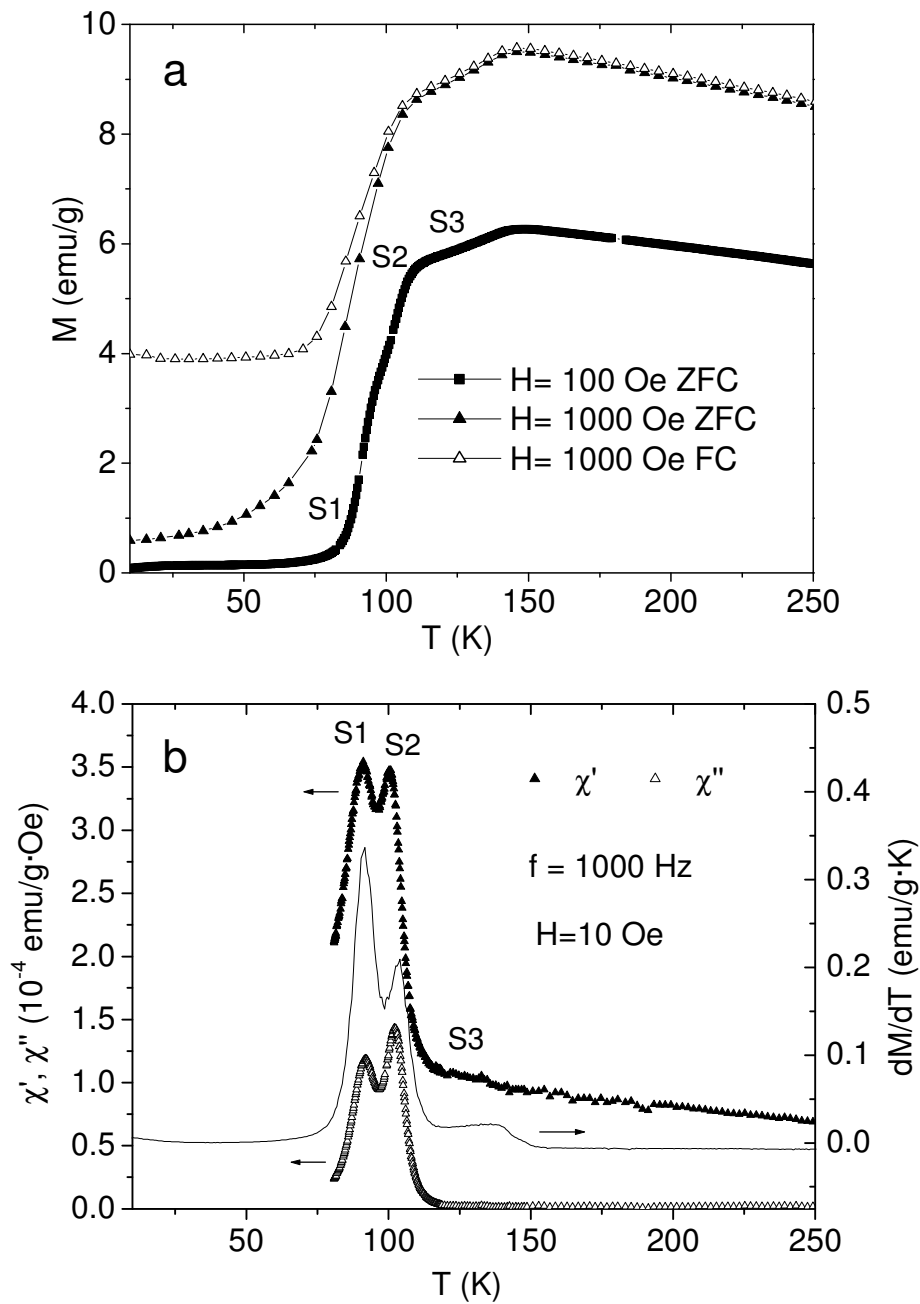

M. Gich *et al.*

**Figure 11**

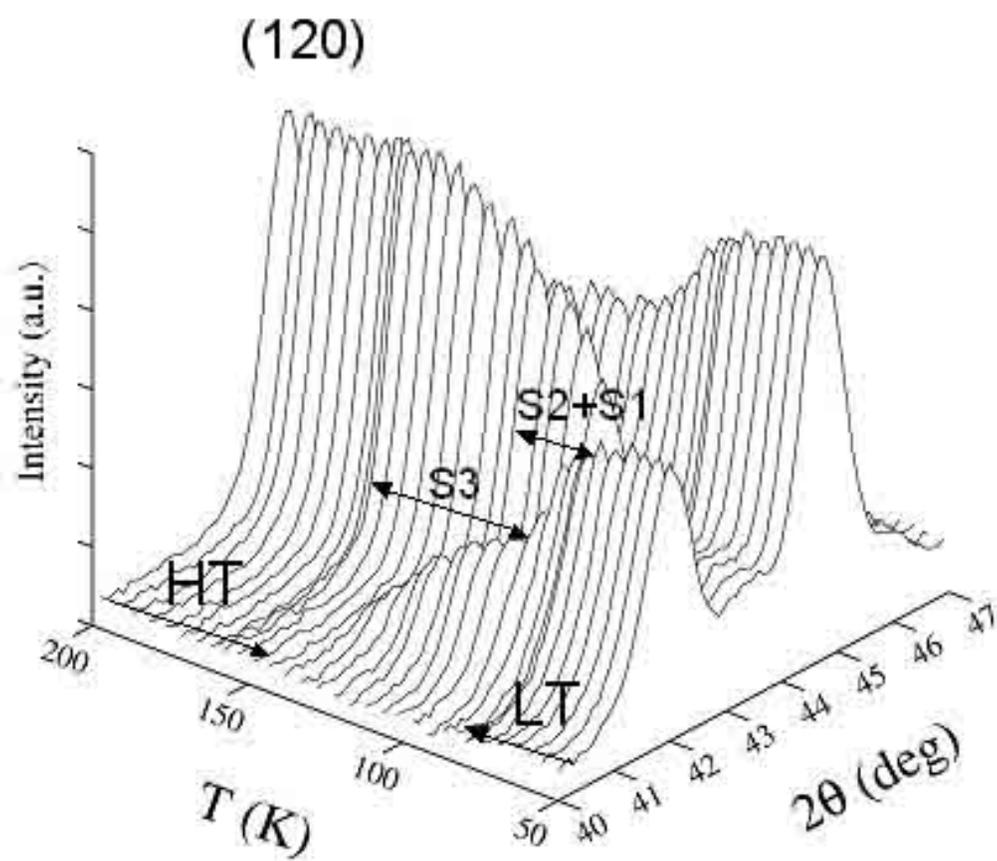

**Figure 12**

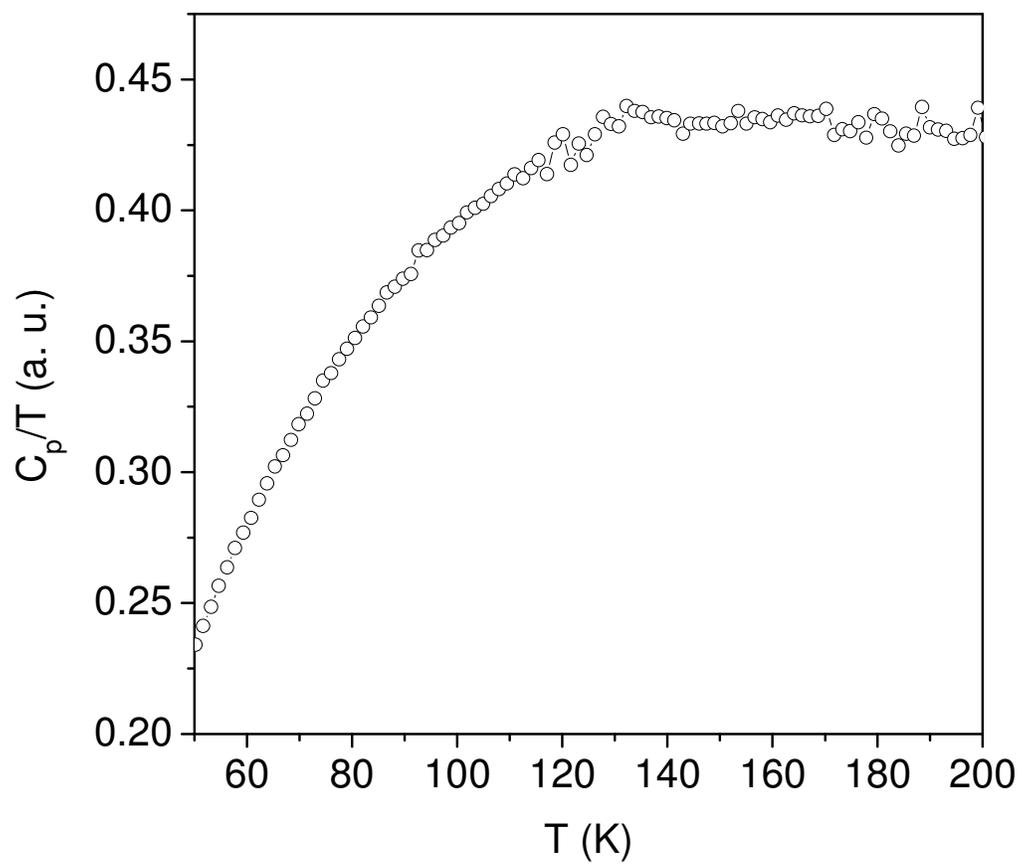

M. Gich *et al.*